\RequirePackage{amsmath}
\documentclass[runningheads]{llncs}
\usepackage{graphicx}
\usepackage[strings]{underscore}
\usepackage{booktabs}
\usepackage[load-configurations=version-1]{siunitx} 
\usepackage{colortbl,etoolbox,xcolor}
\robustify\bfseries
\usepackage{subfig}
\usepackage[misc]{ifsym}

\usepackage{amsmath,amsfonts,mathtools}

\begin{document}
\title{Deep learning-based bias transfer for overcoming laboratory differences of microscopic images}
\titlerunning{Deep learning-based bias transfer for microscopic images}

\author{Ann-Katrin Thebille \inst{1} \and
Esther Dietrich \inst{1} \and
Martin Klaus\inst{1,2} \and
Lukas Gernhold\inst{2} \and
Maximilian Lennartz\inst{3} \and 
Christoph Kuppe\inst{4} \and
Rafael Kramann\inst{4} \and
Tobias B.\ Huber\inst{2} \and
Guido Sauter \inst{3} \and
Victor G.\ Puelles \inst{2} \and
Marina Zimmermann \inst{1,2}\thanks{Joint corresponding/last authors} \textsuperscript{(\Letter)} \and
Stefan Bonn \inst{1}\textsuperscript{$\star$} \textsuperscript{(\Letter)} 
}
\authorrunning{A.-K.\ Thebille et al.}

%
\institute{
Institute of Medical Systems Biology,  Center for Biomedical AI (bAIome), University Medical Center Hamburg-Eppendorf, Hamburg, Germany
\email{marina.zimmermann@zmnh.uni-hamburg.de, sbonn@uke.de}
\and
III. Department of Medicine, University Medical Center Hamburg-Eppendorf, Hamburg, Germany
\and 
Institute of Pathology, University Medical Center Hamburg-Eppendorf, Hamburg, Germany
\and
Institute of Experimental Medicine and Systems Biology, and Division of Nephrology and Clinical Immunology, RWTH Aachen University, Aachen, Germany}
\maketitle              
\begin{abstract}
    The automated analysis of medical images is currently limited by technical and biological noise and bias.
    The same source tissue can be represented by vastly different images if the image acquisition or processing protocols vary.
    For an image analysis pipeline, it is crucial to compensate such biases to avoid misinterpretations.
    Here, we evaluate, compare, and improve existing generative model architectures to overcome domain shifts for immunofluorescence (IF) and Hematoxylin and Eosin (H\&E) stained microscopy images.
    To determine the performance of the generative models, the original and transformed images were segmented or classified by deep neural networks that were trained only on images of the target bias.
    In the scope of our analysis, U-Net cycleGANs trained with an additional identity and an MS-SSIM-based loss and Fixed-Point GANs trained with an additional structure loss led to the best results for the IF and H\&E stained samples, respectively.
    Adapting the bias of the samples significantly improved the pixel-level segmentation for human kidney glomeruli and podocytes and improved the classification accuracy for human prostate biopsies by up to $14\%$.

\end{abstract}

\section{Introduction}
Deep learning (DL) applications play an increasingly important role in medicine, yet, ``everyone participating in medical image evaluation with machine learning is data starved" \cite{Kohli2017}.
When the same imaging technique is used, e.g. confocal microscopy of a specific tissue type with a matching staining protocol, image analysis networks should be applicable to datasets not seen during training without a substantial drop in performance.
However, due to bias introduced during the acquisition or processing of datasets (`domain shift'), the generalizability of deep neural networks is negatively affected.
If bias cannot be accounted for, models have to be re-trained every time a new dataset becomes available.
In consequence, new data can only be used if a large cohort of labeled samples exists or can be created.
Traditional techniques try to compensate domain shift through normalization or simple image transformations, like an increase of the contrast through histogram equalization~\cite{Gonzalez2007}.
However, as indicated by de Bel et al.\ \cite{DeBel2019}, simple approaches are limited in how much bias they can capture and adjust, leading to insufficient transformations.

Recent evidence suggests that deep generative models could be utilized to modify samples, creating novel images that are close to the reference domain while not altering the original content.
This way, images of a new dataset can be adjusted to the bias of the reference domain as a pre-processing step (`bias transfer'), in order to avoid re-training large image analysis networks.
The goal is that new data of the same modality can be handled correctly and without a large drop in performance.
Necessarily, content preservation is of the utmost importance since hallucination artifacts introduced by the generative models could lead to misdiagnosis.
Reliable bias transfer approaches would also enable the usage of DL in settings that do not allow for frequent creation and retraining of models, such as decision support systems in hospitals.  

In this paper, we aim to improve existing generative models to enable stable bias transfer for medical histopathology images.
In addition, we propose guidelines for testing and evaluating bias transfer models in settings with similar transformation goals.
To benchmark the quality of the bias transfer for three state-of-the-art generative models, cycle-consistent generative adversarial networks (cycleGANs) \cite{Zhu2017}, U-Net cycleGANs \cite{DeBel2019}, and Fixed-Point GANs \cite{Choi2018}, we measured the content preservation, target domain adaptation, and impact on image segmentation and classification performance.
To increase the performance of the models, we tested three additional losses designed to improve the quality in terms of content (`MS-SSIM loss')~\cite{Armanious2019}, structure integrity (`structure loss')~\cite{Ma2020} and intensity of transformation (`additional identity loss')~\cite{DeBel2019}.
As a baseline, histogram matching~\cite{Gonzalez2007} for color correction in a decorrelated color space~\cite{Reinhard2001} (`color transfer') is included in our evaluation, which utilizes a single, random image to represent the target domain.

\section{Related work}
\label{sec:relatedwork}
In a medical context, most datasets that require bias transfer are unpaired, meaning that a ground truth for the transformed images does not exist.
A well-established approach for learning unpaired image-to-image translations is using CycleGANs~\cite{Zhu2017}.
CycleGANs have already been used for stain transforming renal tissue sections~\cite{DeBel2019} or histological images of breast cancer~\cite{Shaban2019}.
Unfortunately, every paper introduces a new variation of cycleGAN.
To the best of our knowledge, an extensive comparison of bias transfer algorithms for microscopy images does not exist yet.
Selecting a fitting approach is a non-trivial task.

One common enhancement is using a U-Net structure for the generators~\cite{Manakov2019,DeBel2019}.
In a U-Net, the encoder and decoder structures of the neural network are connected via skip connections between layers of the same size~\cite{Ronneberger2015}.
Thus, the generators can pass information directly without transitioning the bottleneck, improving the level of detail of the images.
In this paper, we refer to the modified cycleGAN approach as U-Net cycleGAN.

Besides architecture modifications or simple hyperparameter adaptations like changing the learning rate~\cite{DeBel2019,Manakov2019}, or the number of images per batch~\cite{DeBel2019,Shaban2019}, a frequent change is adding additional losses to incorporate demands the network has to fulfill.
Armanious et al.\ used the multi-scale structural similarity index (MS-SSIM)~\cite{Wang2003} in \cite{Armanious2019} as an additional cycle loss between the original and the cycle-reconstructed images.
Their goal was to penalize structural discrepancies between the images.
In their experiments, the additional loss led to sharper results with better textural details.
Another loss has been proposed by de Bel et al.\ in~\cite{DeBel2019}.
They included an additional identity loss that is decreased to zero over the first 20 epochs of training.
The loss is an addition to the original identity loss of cycleGANs~\cite{Zhu2017}.
In their experiments, the loss stabilized the training process and led to faster convergence since it forces the generator to look for transformations close to identity first, effectively shrinking the solution space.
Moreover, Ma et al.\ proposed a loss based on the sub-part of the structural similarity index (SSIM) that only evaluates local structure changes~\cite{Ma2020}. 
They proposed the loss to enhance the quality of endoscopy and confocal microscopy images that suffer from ``intensity inhomogeneity, noticeable blur and poor contrast''~\cite{Ma2020}.
The structure loss directly compares the original and transformed images.

Unfortunately, cycleGANs are designed for transforming between two domains only.
Therefore, transforming multiple domains to the reference domain requires individual models and training runs. 
In \cite{Choi2018}, Choi et al.\ proposed StarGANs, which transform between any number of domains with a single generator and discriminator.
The generator produces samples based on the input image and the label of the target domain.
Even if only two domains exist, StarGANs can potentially outperform cycleGAN-based approaches due to the benefit that all data samples are used to train a single generator.
Siddiquee et al.\ proposed Fixed-Point GANs (FPGs), which extend StarGANs with a conditional identity loss, thus reducing hallucination artifacts~\cite{Siddiquee2019}.
If a new dataset could be adjusted by adding a new condition to FPG, bias transfer would be an exceptionally fast solution for the generalizability problem of image analysis networks.

\section{Methodology}
In the following, the overall workflow, as well as the generative models (Subsection \ref{sub:gen}) and our datasets (Subsection \ref{sub:data}) are introduced.
\begin{figure}[!htb]
  \centering
  \includegraphics[width=0.9\textwidth]{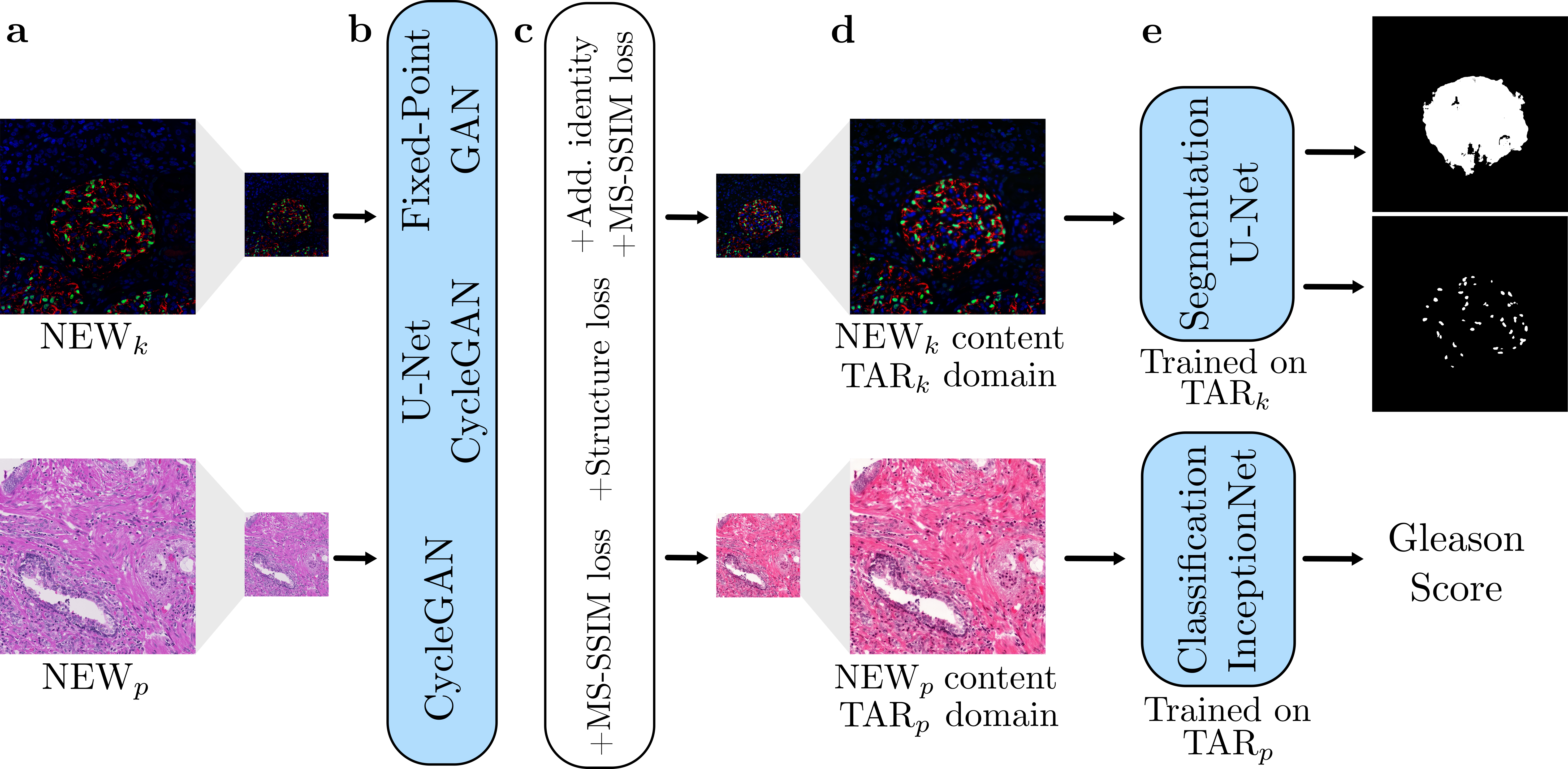}
  \caption{Overview of the image analysis pipeline including bias transfer: First, the original input images are downsampled \textbf{a} and transformed by the generative models \textbf{b} trained with or without additional losses \textbf{c}. Afterwards, they are upsampled to the original size \textbf{d}. The transformed images are then used as inputs for the segmentation or classification networks \textbf{e}. $\textrm{NEW}_{k}$ and $\textrm{TAR}_{k}$ refer to the domains of the kidney samples, whereas $\textrm{NEW}_{p}$ and $\textrm{TAR}_{p}$ are the domains of the prostate samples.
  }
  \label{fig:metho_overview}
\end{figure}

As part of this work, several generative approaches are introduced into an existing image analysis pipeline.
An overview of the complete workflow with bias transfer is shown in Figure~\ref{fig:metho_overview}.
As bias transfer is used as a pre-processing step here, the image size of the transformed images should be identical to the original size.
However, generative models are usually not designed for images larger than $256\times256$ pixels.
We used the Laplacian pyramid-based approach introduced in~\cite{Engin2018} to downsample the images to $256\times256$ pixels pre- (Figure~\ref{fig:metho_overview} a $\rightarrow$ b), and back to the original size post-transformation (Figure~\ref{fig:metho_overview} c $\rightarrow$ d), which is $1024\times1024$ pixels for the kidney (upper row) and $2048\times2048$ for the prostate images (lower row).
The approach involves replacing the smallest layer of the pyramids with the respective transformed image for upsampling.
Hence, the edges of the original image, which are typically lost at a low resolution, are added back into the image, boosting the quality of the transformed image.

\subsection{Generative approaches}\label{sub:gen}
For this paper, we implemented, tested, and modified three state-of-the-art generative models for image generation, as shown in Figure~\ref{fig:metho_overview}b.
For this purpose, we re-created the original implementation by Zhu et al.\ for cycleGANs~\cite{Zhu2017} in Tensorflow 2.0 with a PatchGAN discriminator that judges $16\times16$ patches per image.
The U-Net cycleGAN architecture has been constructed by adding skip connections to the generators. 

For FPG, the generator input consists of an image and the one-hot encoded label of the target domain.
Otherwise the generator architecture is identical to cycleGAN.
Instead of using the discriminator structure described in the original FPG paper, we created a modified version of the cycleGAN discriminator by replacing the original output layer with two outputs, one for predicting the image's domain and one for the authenticity of the individual image patches (real vs.\ generated).
The modified cycleGAN discriminator has fewer trainable parameters and judges more patches per image than the original FPG discriminator. 
In~\cite{Isola2017} it has been shown that judging more patches per image leads to images with more detail for cycleGANs.
It seems evident that FPGs could also benefit from this modification.

The hyperparameters for cycleGAN and U-Net cycleGAN are mainly based on the original cycleGAN implementation.
The networks are trained for up to 200 epochs with an initial steady learning rate of $0.0005$ (original: $0.0002$) for the first 100 epochs and a linearly decaying learning rate that reaches 0 after 200 epochs.
For FPG, an initial learning rate of $0.0001$ has been used, as proposed by Choi et al.\ \cite{Choi2018}.
We used a batch size of one.

To improve the generative performance and reduce hallucination artifacts of the three models, we included additional terms in the loss functions (see Figure~\ref{fig:metho_overview}c) specifically the additional identity loss $\mathcal{L}_{\textrm{+id}}$~\cite{DeBel2019}, the MS-SSIM loss $\mathcal{L}_{\textrm{+ms-ssim}}$~\cite{Armanious2019} and the structure loss $\mathcal{L}_{\textrm{+struc}}$~\cite{Ma2020}.
The original losses of (U-Net) cycleGANs are the adversarial loss ($adv$), the cycle-reconstruction loss ($cyc$) and the identity loss ($id$).
For FPGs, the original losses are the cycle-reconstruction loss ($cyc$), the domain-classification loss ($domain$), the gradient penalty loss ($gp$) and the conditional identity loss ($id$).
For our experiments, the losses have been weighted with $\lambda_{\textrm{adv}} =1$, $\lambda_{\textrm{cyc}} =10$, and $\lambda_{\textrm{id}}=10$ (original: $\lambda_{\textrm{id}}=5$) for cycleGAN and U-Net cycleGAN and with  $\lambda_{\textrm{cyc}}=\lambda_{\textrm{gp}}= \lambda_{\textrm{id}} = 10$ and $\lambda_{\textrm{domain}}=1$ for FPG.
All additional losses have been weighted with $\lambda=5$.
We added the structure loss and the MS-SSIM loss individually and, for the MS-SSIM loss, in combination with the additional identity loss (`combined losses').
The structure loss was not combined with the other additional losses since the direct comparison of the original and transformed images covers similar objectives as the `combined losses'.
Our implementation is publicly available on GitHub.\footnote{\url{https://github.com/imsb-uke/bias-transfer-microscopy}}

\subsection{Data}\label{sub:data}
In this paper, we perform bias transfer for two modalities, IF images of kidney biopsies and H\&E stained tissue microarray (TMA) spots of prostate biopsies.
In Figure~\ref{fig:metho_overview}, the upper row shows the workflow for the kidney and the lower row for the prostate biopsy samples.
Each modality includes two sub-datasets (domains) originating from different hospitals and has a specified transformation direction.
This is due to the fact that bias transfer is applied here to overcome the domain shift between a new and a target domain which has been used for training a segmentation or classification network, as outlined in Figure~\ref{fig:metho_overview}e.

\subsubsection{Kidney biopsies}
The kidney dataset consists of 2D kidney IF images~\cite{Zimmermann2021}.
The images have been used to train a modified U-Net with dual output for the automatic segmentation of the glomeruli and their podocytes, an important cell type inside the glomeruli that is indicative of the health of the kidney~\cite{Zimmermann2021}.
To highlight the glomeruli and podocytes, three different biomarkers were used: DAPI for cell nuclei (blue), WT1 for podocyte cytoplasm (red), and DACH1 for podocyte nuclei (green).
The images originate from two distinct hospitals (two domains: the target domain `$\textrm{TAR}_{k}$' and the new domain `$\textrm{NEW}_{k}$'), which were imaged by two different operators with differing confocal laser scanning microscopes (Nikon A1 Confocal and Zeiss Confocal LSM 800).
In total, subset $\textrm{TAR}_{k}$ contains 269 images and subset $\textrm{NEW}_{k}$ contains 375 images.
Training and evaluating the segmentation requires annotations.
For this purpose, a subset of the images has been annotated by medical experts ($\textrm{TAR}_{k}$: 109, $\textrm{NEW}_{k}$: 90).
Further details on the datasets and their biological meaning can be found in~\cite{Zimmermann2021}.

For bias transfer, the split into training, validation, and test sets was performed randomly, with the constraints that images originating from the same patient (up to 16) remain in the same set, that a ratio of approximately 70\% for training and 15\% each for validation and test sets is reached and that the validation and test sets only consist of annotated images.
While bias transfer does not require a ground truth, masks are necessary to evaluate the segmentation quality achieved on the transformed images.
Accordingly, the $\textrm{TAR}_{k}$ images were split into 180 for training, 44 for validation, and 45 for testing.
The $\textrm{NEW}_{k}$ dataset was split into 285 images for training, 46 for validation, and 44 for testing.

\subsubsection{Prostate biopsies}
The prostate dataset consists of circular biopsies (spots) originating from radical prostatectomies (RPE) that have been assembled with the help of TMAs.
The images originate from two different hospitals (`$\textrm{TAR}_{p}$' and `$\textrm{NEW}_{p}$'). 
The $\textrm{TAR}_{p}$ dataset consists of 2866 and the $\textrm{NEW}_{p}$ dataset of 886 images.
Both datasets were created for staging prostate cancer with the help of Gleason patterns, which categorize the shape of the glands.
Tissue can be classified as benign or as Gleason pattern 3, 4, or 5, whereas a higher number represents worse tumor tissue.
The Gleason score (GS) is then calculated as the sum of the most prevalent and the worst occurring patterns in a sample~\cite{Chen2016}.
Unfortunately, the inter-pathologist agreement on GSs is usually low~\cite{Egevad2013}, which is why automated and stable GS prediction is of high clinical value.

We trained an InceptionNet-V3-based classification network, pre-trained on ImageNet~\cite{Deng2009}, on images of the $\textrm{TAR}_{p}$ dataset, reaching a test accuracy of $81.6\%$.
To exclude image background only the innermost $2048\times2048$ pixels of the spots have been used. 
The classification was limited to single Gleason patterns only (`0' = benign, `3+3', `4+4', and `5+5') since TMA spots with two (or more) differing Gleason patterns could potentially contain patterns that occur solely outside of the innermost pixels.
During training, all images are pre-processed with normalization, and the data is augmented with random rotations, shearing, shifting, and flipping. 
After pre-processing, the images are resized to $224\times224$ pixels, which is the input size of the InceptionNet.

The $\textrm{NEW}_{p}$ dataset is publicly available~\cite{Arvaniti2018a} and does not include Gleason score annotations.
However, the images have been annotated via segmentation masks to identify areas containing a specific Gleason pattern.
To evaluate the classification performance of our network, we used the segmentation masks to calculate the Gleason scores of the image centers.
We then used the same split into training, validation, and test sets that was defined for the dataset in the original publication~\cite{Arvaniti2018b}.
The training and validation sets have been annotated by one pathologist (pathologist 1) and the test set has been annotated independently by two pathologists (pathologist 1 and pathologist 2), resulting in 506 images (429 with a single Gleason pattern) for training, 133 (111) for validation, and 245 (199 -- pathologist 1, 156 -- pathologist 2) images for testing.
For the $\textrm{TAR}_{p}$ dataset, the split was determined according to a random stratified sampling strategy, resulting in the same label distribution for the training, validation, and test sets.
The resulting split consists of 2000 (1136 with a single Gleason pattern) images for training, 432 (246) images for validation, and 434 (245) images for testing.
All images were annotated by the same pathologist (pathologist 3).
Nonetheless, the label distribution is not uniform, $\textrm{TAR}_{p}$ overrepresents 3+3 samples. 
In contrast, $\textrm{NEW}_{p}$ mostly contains 4+4 samples.

\section{Results and discussion}
We based the evaluation of the generative approaches on three factors: content preservation, target domain imitation, and impact on the segmentation or classification performance.
Two metrics have been selected for a quantitative evaluation of the transformation quality: the structural similarity index (SSIM)~\cite{Wang2004} and the Fr\'echet Inception Distance (FID)~\cite{Heusel2017}.
The SSIM is a well-established metric that has been used here to calculate the degradation of structural information (content) between the original (Figure~\ref{fig:metho_overview}a) and the transformed (Figure~\ref{fig:metho_overview}d) input images.
The FID measures the feature-wise distance between two domains.
While bias transfer can be trained on all images, for the prostate dataset, the impact on the classification can only be evaluated on the single Gleason pattern images.
Nonetheless, we decided to evaluate all images regarding the SSIM and FID since the differing annotations given by pathologist 1 and 2 create two non-identical subsets of test images.
For our evaluation, the FID between the target domain and the transformed images is calculated and compared to the FID between the original domains.
A lower FID implies that images are visually closer to the target domain after bias transfer.
Since the FID highly depends on the image content, validation and test scores should be considered separately.

Finally, for the kidney data, the segmentation accuracy pre- and post-bias transfer is compared.
The segmentation network predicts one segmentation mask for the glomerulus as a whole and one for the podocytes (see Figure~\ref{fig:metho_overview}e).
The quality of the segmentation is measured with three Dice scores~\cite{Dice1945}, the pixel-wise segmentation of the glomeruli and podocytes and the object-wise segmentation of the podocytes.
For the prostate data, the classification accuracy and the macro F1 scores pre- and post-bias transfer are compared.
The macro F1 score can give insights into the classification performance for underrepresented classes.
Considering class imbalance is especially important here since the label distributions of $\textrm{TAR}_{p}$ and $\textrm{NEW}_{p}$ do not match. 

Every variation has been trained five times with different random seeds, including the baseline.
The training epoch with the lowest generator validation loss for transformations from $\textrm{NEW}$ to $\textrm{TAR}$ has been used for the evaluation of the individual runs.
For each approach, the run with the best results on the validation set has been evaluated on the test set.

\subsection{Results}
\subsubsection{Kidney biopsies}
\begin{figure}[p]
  \centering
  {
    \subfloat[$1-\textrm{SSIM}$]{%
      \includegraphics[width=0.47\textwidth]{{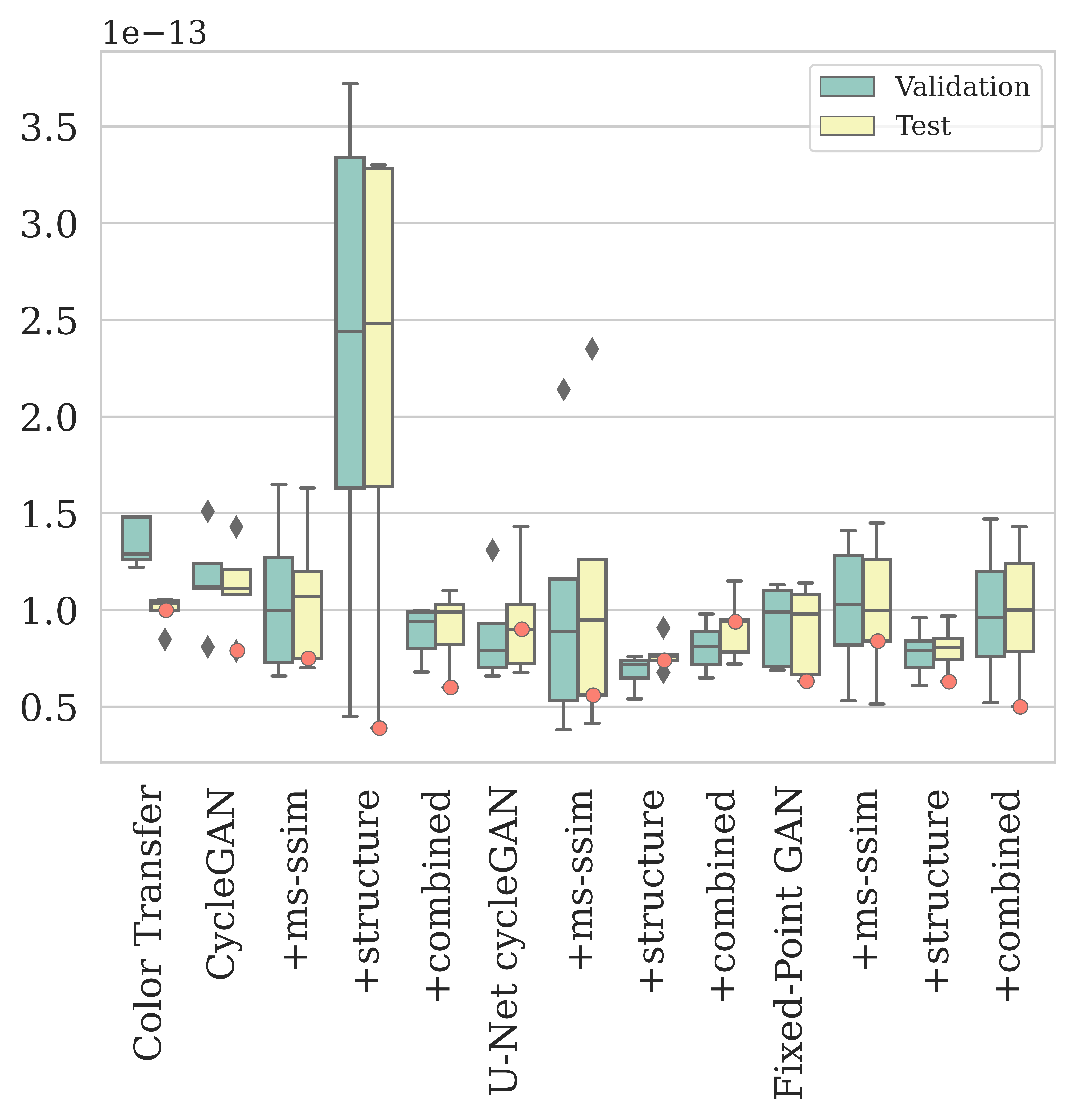}}
      \label{fig:ssim_boxplot_kidney}
    }
    \subfloat[$\textrm{FID}$]{
      \includegraphics[width=0.47\textwidth]{{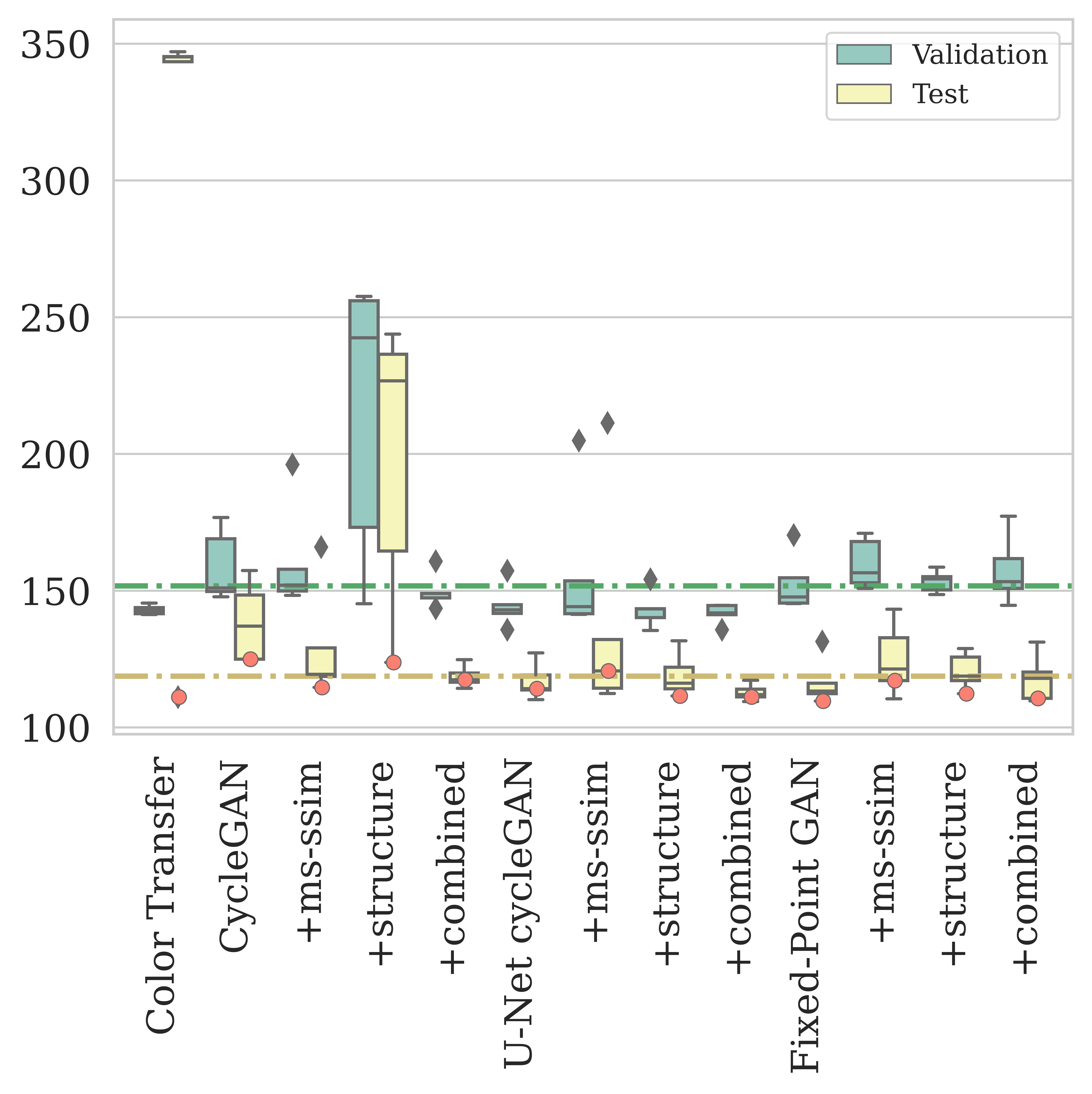}}
      \label{fig:fid_boxplot_kidney}
      }
  }
  \caption{Boxplots visualizing the transformation metrics $1-\textrm{SSIM}$ (a) and FID (b) for the validation and test sets of the kidney biopsies. The dashed lines highlight the original FID scores and the red dots show the values achieved by the runs that performed best on the validation set for each variation.}
  \label{fig:boxplots_kidney}
\end{figure}

The SSIM and FID scores for the validation set are visualized in Figure~\ref{fig:boxplots_kidney}.
They indicate that the transformations performed by the U-Net cycleGAN and FPG variations had the highest quality, with U-Net cycleGAN with structure loss leading to the overall best and most stable SSIM scores.
Regarding the FID, U-Net cycleGAN with combined losses largely improved the distance to the target domain for the validation and test sets.
The combination of additional identity and MS-SSIM loss (combined losses) had a stabilizing effect on the training process of cycleGAN and U-Net cycleGAN.
The structure loss had a positive effect on U-Net cycleGAN and FPG, however, it led to mode collapse for cycleGAN, only producing a single output irrespective of the input.

\begin{table}[p]
    \centering
    \caption{Means ($\mu_{test}$) and standard deviations ($\sigma_{test}$) of the Dice scores on the original test sets and the relative performance for the transformed images. `Dice glom.\ pix.' evaluates the pixel-wise segmentation of the glomeruli, and `Dice podo.\ pix.' of the podocytes. `Dice podo.\ obj.' evaluates the object-wise segmentation of the podocytes. Significance ($p<0.05$) is marked with $*$.}
    \label{tab:kidney_segmentation_dice}
    \begin{tabular}{lS[retain-explicit-plus, table-format=-1.3, table-space-text-post={*} ]S[table-format=1.2]S[retain-explicit-plus,table-format=-1.3, table-space-text-post={*}]S[table-format=1.2]S[retain-explicit-plus, table-format=-1.3, table-space-text-post={*}]S[table-format=1.2]}
    \toprule
        & \multicolumn{2}{S}{\bfseries{Dice glom. pix.}} & \multicolumn{2}{S}{\bfseries{Dice podo.\ obj.}} & \multicolumn{2}{S}{\bfseries{Dice podo.\ pix.}} \\
    \cmidrule(lr){2-3}
    \cmidrule(lr){4-5}
    \cmidrule(lr){6-7}
            & {$\mu_{test}$} &  {$\sigma_{test}$} & {$\mu_{test}$} &  {$\sigma_{test}$} & {$\mu_{test}$} & {$\sigma_{test}$}  \\
   \midrule
    Test $\textrm{TAR}_{k}$  & 0.953  & 0.02 & 0.933 & 0.06 & 0.929 & 0.01 \\
    Test $\textrm{NEW}_{k}$  &  0.909 & 0.05 & 0.877 &0.05 & 0.730 & 0.07 \\
    \midrule
    Color transfer & -0.158 * & 0.32 & -0.163 * & 0.27 & -0.079 & 0.26 \\
     \midrule
     CycleGAN                   & -0.002  & 0.06 & -0.053 *    & 0.10 & +0.020  & 0.05\\
     $\ $ + MS-SSIM        & +0.014 &  0.05 & -0.013  & 0.10 & +0.055  * & 0.04 \\
     $\ $ + structure & -0.020  & 0.10 &  -0.055* & 0.09 & -0.017  & 0.09\\
     $\ $ + combined    & -0.005  *    & 0.05 & -0.031 *    & 0.09 & +0.046 & 0.05\\
     U-Net CycleGAN             & +0.018 *          & 0.04 & \bfseries +0.015 & 0.05 & +0.061  * & 0.04 \\
     $\ $ + MS-SSIM      & +0.020 *          & 0.03 & -0.002     & 0.07 & +0.039* & 0.04 \\
     $\ $ + structure & +0.017 & 0.04 & -0.023 & 0.11 & +0.064* & 0.05 \\
     $\ $ + combined    & +0.022*          & 0.03 &  +0.000 & 0.08 & \bfseries +0.067* & 0.03 \\
     Fixed-Point GAN            & -0.002  & 0.06 & -0.019 & 0.09 & +0.060* & 0.05 \\
     $\ $ + MS-SSIM        & \bfseries +0.031* & 0.02 & -0.005 & 0.07 & +0.040* & 0.04\\
     $\ $ + structure & +0.022* & 0.04 & -0.007 & 0.07 & +0.055* & 0.04 \\ 
     $\ $ + combined    & +0.025* &    0.03 & -0.002& 0.07 & +0.048* & 0.04 \\
    \bottomrule
  \end{tabular}
\end{table}

The segmentation scores on the test set are shown in Table \ref{tab:kidney_segmentation_dice}.
Here, `Test $\textrm{TAR}_{k}$' only includes images not used for training the segmentation U-Net (21 images).
Color transfer worsened the segmentation scores, despite improving the FID.
As for the SSIM and FID scores, the U-Net cycleGAN and Fixed-Point GAN variations led to the best results.
For those approaches, three out of four variations significantly improved the pixel-level Dice score for the glomeruli and all four for the podocytes.
However, unlike FPG, not all runs of U-Net cycleGAN worsened the object-level Dice score for the podocytes.
Overall, U-Net cycleGANs with combined losses performed the best for the task at hand.
The aforementioned approach produced hallucination artifact-free images that significantly improved the pixel-level segmentation scores of the glomeruli (0.909 to 0.923, $p=0.005$), and podocytes (0.730 to 0.797, $p<0.0001$), due to a strong adaption to the target domain.
An example transformation performed by U-Net cycleGAN and its effect on the segmentation result can be found in~\cite{Zimmermann2021}.

\subsubsection{Prostate biopsies}
\begin{figure}[p]
  \centering
  {
    \subfloat[$1-\textrm{SSIM}$]{%
      \includegraphics[width=0.47\textwidth]{{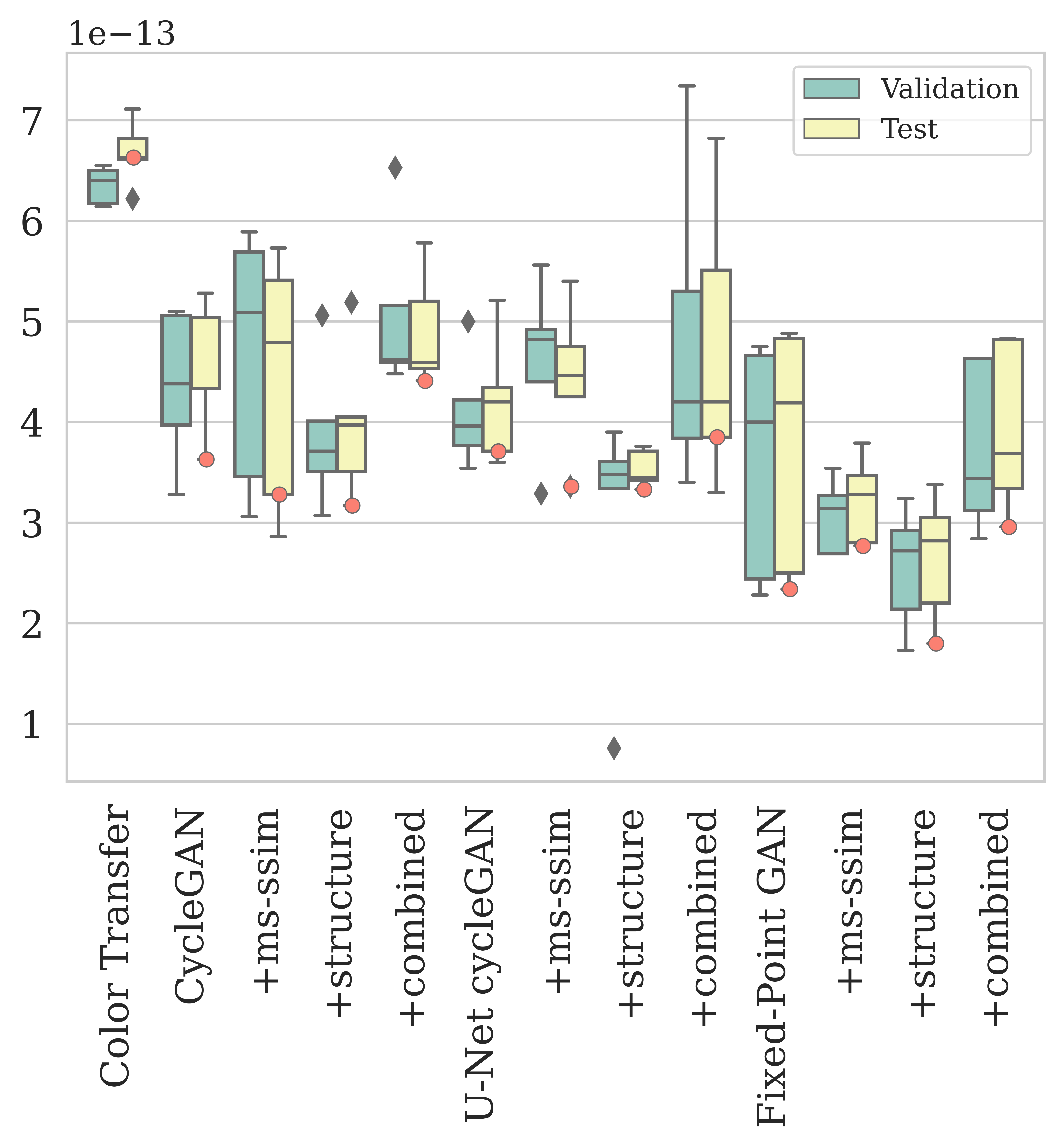}}
      \label{fig:ssim_boxplot_prostate}
      }
    \subfloat[$\textrm{FID}$]{
      \includegraphics[width=0.47\textwidth]{{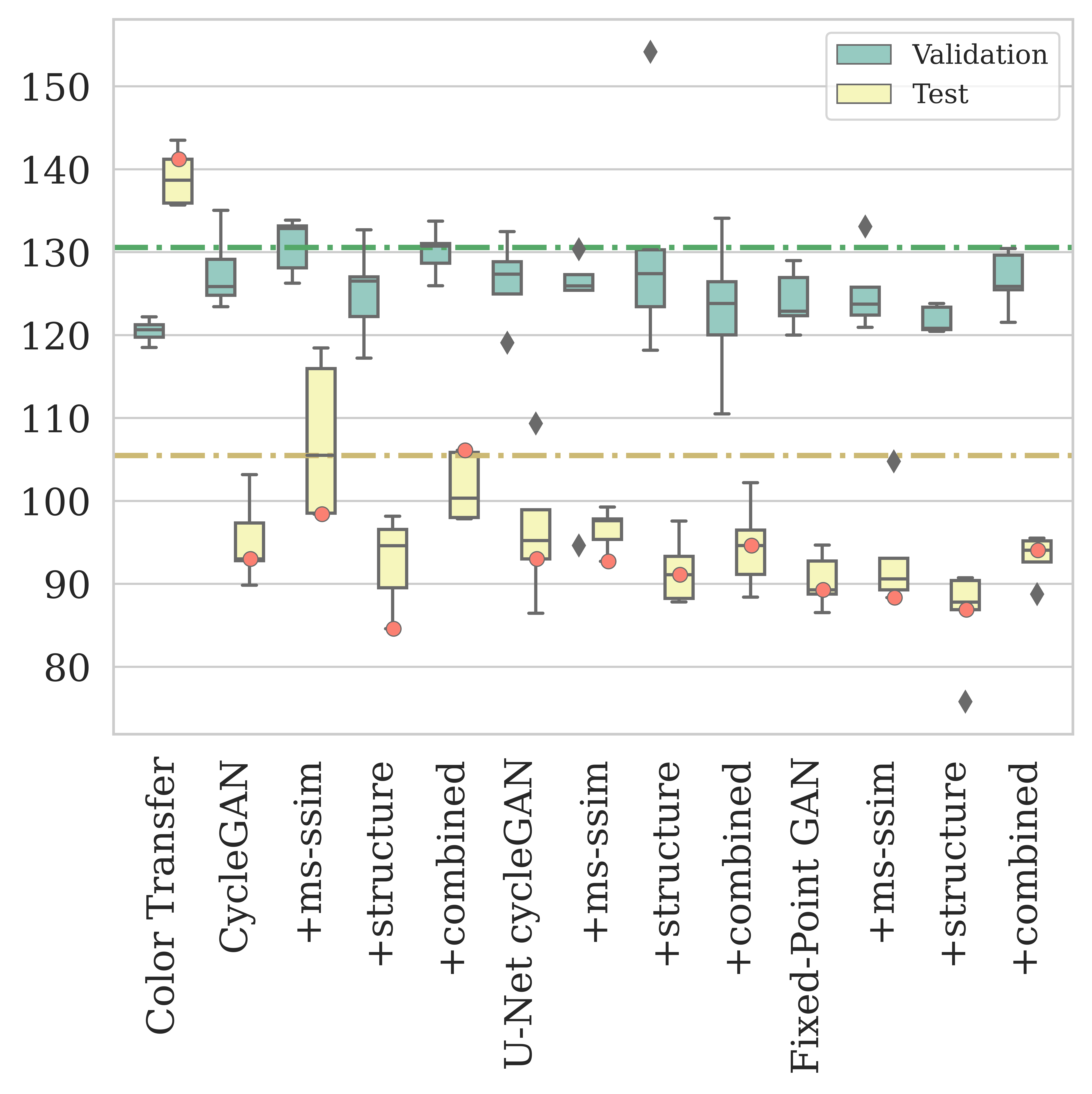}}
      \label{fig:fid_boxplot_prostate}
      }
  }
  \caption{Boxplots visualizing the transformation metrics $1-\textrm{SSIM}$ (a) and FID (b) for the validation and test sets of the prostate biopsies. The dashed lines highlight the original FID scores and the red dots mark the results achieved by the runs that performed best on the validation set for each variation.}
  \label{fig:boxplots_prostate}
\end{figure}

For the prostate biopsies, the boxplots in Figure~\ref{fig:boxplots_prostate} show that FPG trained with the structure loss outperformed the other approaches regarding content preservation.
While the overall lowest FID score on the test set was achieved by cycleGAN with structure loss, this result is not reproducible for different random seeds -- the training process is not stable enough.

\begin{table}[p]
  \centering
  \caption{Means ($\mu_{val}$) and standard deviations ($\sigma_{val}$) of the accuracy and macro-weighted F1 scores for the classification of the Gleason scores on the original validation and test sets and the relative performance for the transformed images. For the test set, the predicted Gleason scores are compared to the annotations of two medical professionals ($\mu_{test 1}$ and $\mu_{test 2}$). The best results are bold.}
  \label{tab:prostate_classification_metrics}
  \begin{tabular}{lS[retain-explicit-plus, table-format=-1.3]S[table-format=1.2]S[retain-explicit-plus,table-format=-1.3]S[retain-explicit-plus,table-format=-1.3]S[retain-explicit-plus, table-format=-1.3]S[table-format=1.2]S[retain-explicit-plus,table-format=-1.3]S[retain-explicit-plus,table-format=-1.3]}
    \toprule
    & \multicolumn{4}{S}{\bfseries{Accuracy}} & \multicolumn{4}{S}{\bfseries{F1 score}} \\
    \cmidrule(lr){2-5}
    \cmidrule(lr){6-9}
    & {$\mu_{val}$} & {$\sigma_{val}$} & {$\mu_{test 1}$} & {$\mu_{test 2}$} & {$\mu_{val}$} & {$\sigma_{val}$} & {$\mu_{test 1}$} & {$\mu_{test 2}$} \\
    \midrule
    Test $\textrm{TAR}_{p}$  & 0.763 & 0.00 & 0.816 & 0.816 & 0.665 & 0.00 & 0.742 & 0.742 \\
    Test $\textrm{NEW}_{p}$  & 0.576 & 0.00 & 0.256 & 0.217 & 0.454 & 0.00 & 0.249 & 0.242 \\
    \midrule
    Color transfer & -0.117 & 0.01 & -0.105 &  -0.108 & -0.178 & 0.03 & -0.096  & -0.107 \\
    \midrule
     CycleGAN                   & +0.073 & 0.04 & +0.090 & +0.038 & +0.066 & 0.04 & +0.069 & +0.032 \\
     $\ $ + MS-SSIM         & +0.063 & 0.06 &  +0.065 & +0.012 & +0.086 & 0.04 & +0.050  & +0.009 \\
     $\ $ + structure               & +0.057 & 0.03 & +0.100 & +0.083 & +0.046 & 0.03 & +0.064  &  +0.060  \\
     $\ $ + combined      & +0.111 & 0.00 & +0.140 & \bfseries +0.141 & \bfseries +0.120 & 0.01 & +0.108  & \bfseries +0.128 \\
     U-Net CycleGAN              & +0.095 & 0.03 & +0.095  & +0.070 & +0.086 & 0.02 & +0.072 & +0.066 \\
     $\ $ + MS-SSIM         & +0.099 & 0.02 & +0.135 & \bfseries +0.141 & +0.102 & 0.02 & +0.107  & +0.120 \\
     $\ $ + structure             & +0.097 & 0.04 & +0.100 & +0.096 & +0.090 & 0.03 & +0.075  & +0.086 \\
     $\ $ + combined     & +0.064 & 0.03 & +0.120 & +0.083 & +0.065 & 0.03 & +0.082  & +0.067 \\
     Fixed-Point GAN             & +0.109 &  +0.00 & \bfseries +0.145  & +0.128 & +0.106 & 0.01 & \bfseries+0.112 & +0.113 \\
     $\ $ + MS-SSIM       & +0.117 & 0.00 & +0.090 & +0.051  & +0.106 & 0.00 & +0.068 &  +0.054 \\
     $\ $ + structure            & \bfseries+0.122 & 0.02 & +0.125 & +0.089 & +0.107 & 0.02 & +0.097  & +0.082 \\
     $\ $ + combined      & +0.117 & 0.01 & +0.140 & +0.096 & +0.109 & 0.00 & +0.111  &  +0.086 \\
    \bottomrule
  \end{tabular}
\end{table}

\begin{figure}[tb]
  \centering
  {
    \subfloat[Confusion matrices]{
      \includegraphics[width=0.31\textwidth]{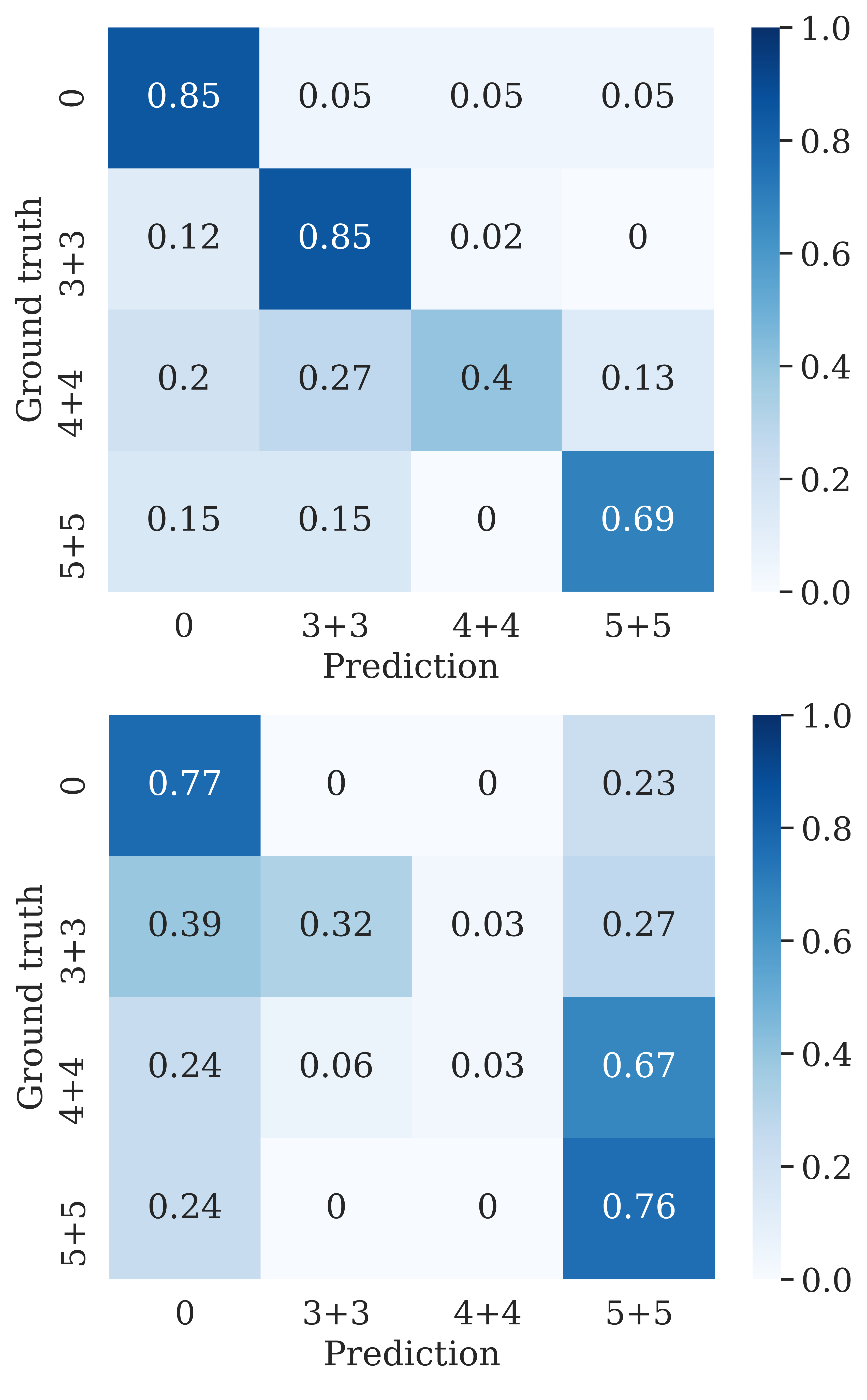}
      \label{fig:tarp_newp1_test_confusion}
      }
    \subfloat[FPG + structure]{%
      \includegraphics[width=0.31\textwidth]{{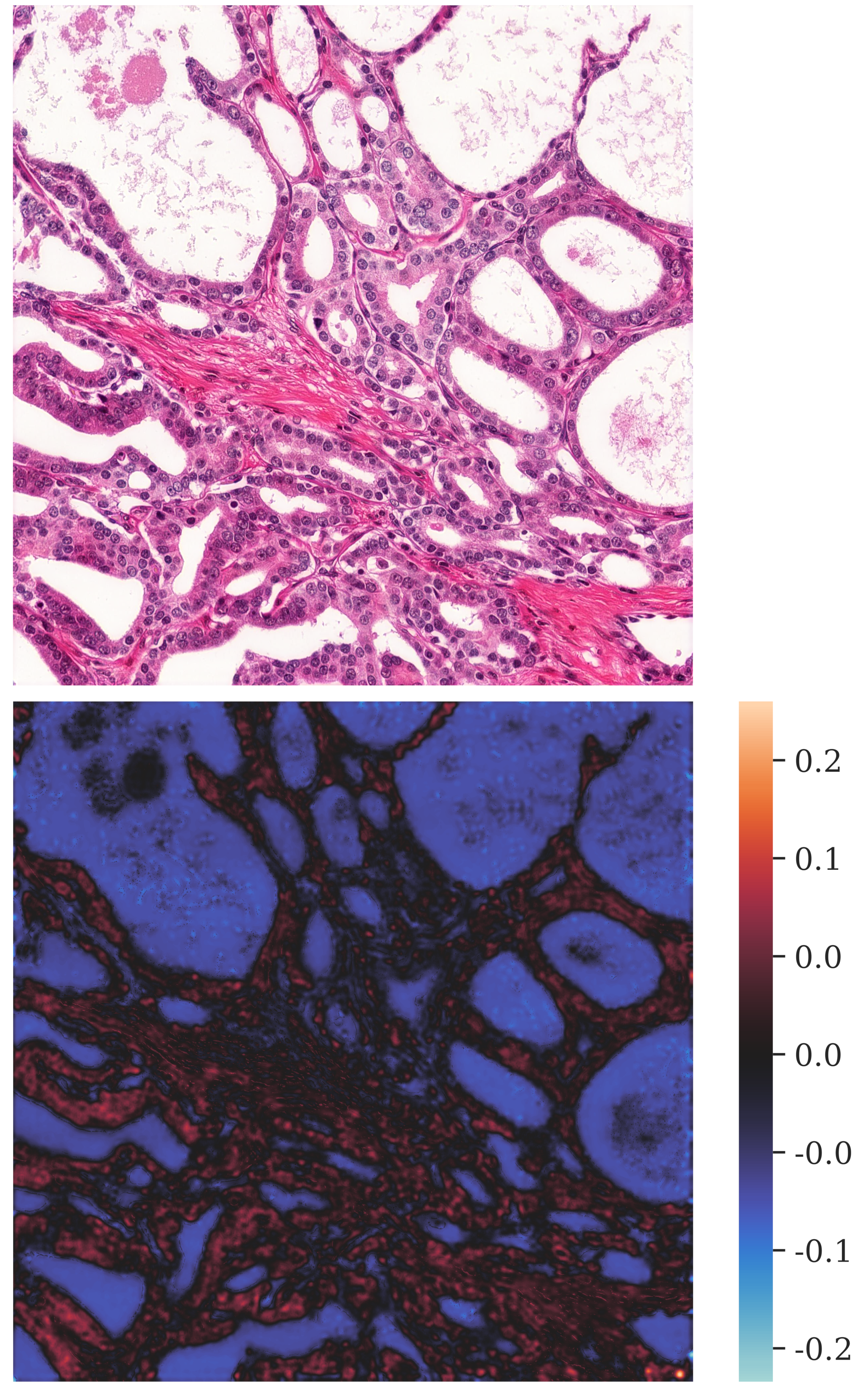}}
      \label{fig:fpg_ma_image}
      }%
    \subfloat[CycleGAN + comb.]{
      \includegraphics[width=0.31\textwidth]{{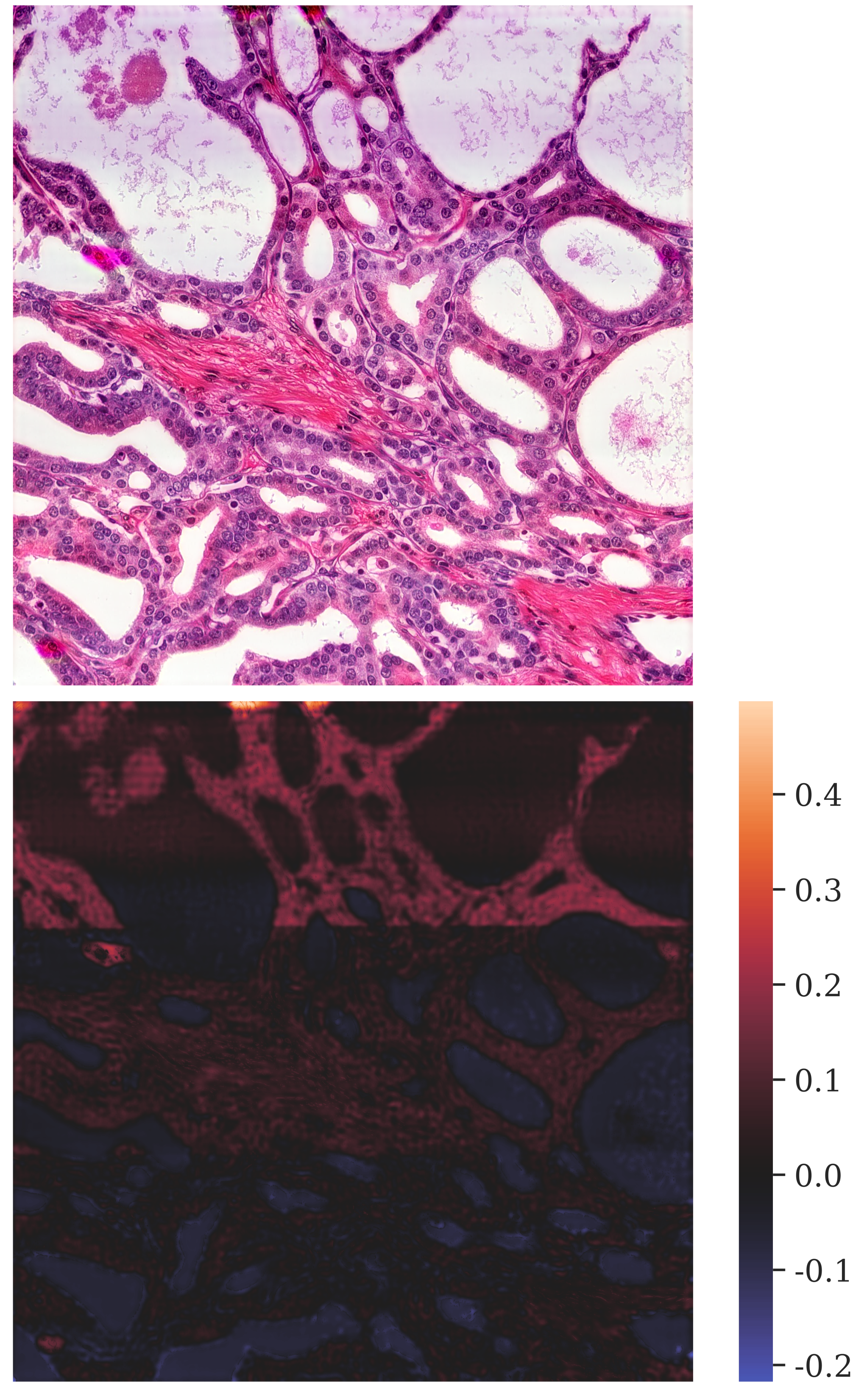}}
      \label{fig:cyc_com_image}
      }
  }
  \caption{(a) Relative confusion matrices for the classification of the test sets for $\textrm{TAR}_{p}$ (top) and $\textrm{NEW}_{p}$ annotated by pathologist 1 (bottom), as well as transformed test images with and without hallucination artifacts and their corresponding transformation heatmaps in (b) and (c).}
  \label{fig:artifact_examples_prostate}
\end{figure}

When we look at the accuracies achieved on the validation set in Table~\ref{tab:prostate_classification_metrics}, FPG with structure loss achieved the largest improvement (from $0.576$ to $0.698$).
However, this was not reproduced on the test set or for the macro-weighted F1 scores.
Regarding the test accuracies, cycleGAN with combined losses and FPG resulted in the best scores.
While FPG achieved the best accuracy and F1 scores for the images annotated by pathologist 1, cycleGAN with combined losses outperformed all other approaches for the images annotated by pathologist 2.
Yet, the images transformed by cycleGAN with combined losses contained hallucination artifacts, as is reflected in the SSIM scores (Figure~\ref{fig:ssim_boxplot_prostate}).
Unfortunately, for the classification network at hand, a direct link between transformation and classification quality does not seem to exist.
In Figure~\ref{fig:cyc_com_image} an image transformed by cycleGAN with combined losses can be seen.
The transformation adds a purple streak across the upper third of the image, which is also reflected in the heatmap of the changes in pixel intensities through the transformation.
In contrast, the change heatmap of the images transformed by FPG with structure loss indicates that the color intensity of the background was reduced and that small cell structures remained intact (see Figure~\ref{fig:fpg_ma_image}).
Hence, to avoid basing the classification on hallucination artifacts, the best bias-transfer approach cannot be judged based on classification accuracies only.

Another factor that could contribute to the mismatch of transformation and classification quality is the very low accuracy on the original $\textrm{NEW}_{p}$ test images, which is $25.6\%$ for the images annotated by pathologist 1 and $21.7\%$ for pathologist 2, compared to $81.6\%$ achieved on the $\textrm{TAR}_{p}$ test set.
The samples of the two domains were annotated by different pathologists, who might interpret structures disparately, resulting in different Gleason scores.
As a result, a transformation to the target domain might not be sufficient to reach high accuracy on this dataset.
This is amplified by the fact that the classification network has the lowest accuracy for 4+4 samples (see Figure~\ref{fig:tarp_newp1_test_confusion}), which is the most frequent class in $\textrm{NEW}_{p}$.
For the original images of $\textrm{NEW}_{p}$ most of the 4+4 samples were misclassified as 5+5 or 0 (benign) (see Figure~\ref{fig:tarp_newp1_test_confusion}).
As a consequence, an accidental improvement of the classification accuracy due to hallucination artifacts cannot be ruled out.
All those reasons indicate that FPG with structure loss may still be considered the best performing bias transfer for the prostate biopsy samples, despite not resulting in the highest classification scores.
It stably led to improved accuracies and F1 scores across all runs while keeping the balance between not adding hallucination artifacts to the images and imitating the target domain.

\subsection{Discussion}
%
%
Currently, bias can be overcome with hand-engineered methods~\cite{DeBel2019} or with deep generative models, which are universal function approximators.
While our datasets do not require a strong transformation, the simple baseline approach was not a sufficient solution.
Like for many simple approaches, the transformation is applied uniformly across the whole image. 
Therefore, non-linear bias cannot be captured.
The resulting, tinted images hindered the segmentation of the glomeruli and their podocytes, as well as the classification of prostate cancer.
In contrast, nine generative approaches significantly improved the pixel-wise segmentation of the podocytes and six the segmentation of the glomeruli.
The classification accuracies and F1 scores were improved by every variation.
However, as indicated by the varying SSIM scores, not all approaches resulted in artifact-free images.
In our experiments, the widely used cycleGAN architecture did not perform well for bias transfer without any modifications. 
The training process was unstable and often led to hallucination artifacts.
This matches reports in previous studies. 
Manakov et al.\ described that, in their experiments, vanilla cycleGANs generated blurry images with checkerboard-like artifacts~\cite{Manakov2019}.

CycleGANs were created for general image-to-image transformation.
While goals like identity-transforming images that belong to the target domain and reconstructing the original from a transformed image are already incorporated in the basic model, other objectives have to be added explicitly and balanced carefully to achieve maximum performance for the task at hand.
Otherwise the instability of the cycleGAN training process can be amplified, leading to complete mode collapse, as we experienced with the kidney biopsies.

FPGs were invented to convert diseased to healthy samples, which requires removing structures from the images.
For bias transfer, however, the content of the image has to stay the same.
Since this is not explicitly safeguarded by FPG, it is not surprising that the additional losses had a positive impact on the transformation.
U-Net cycleGANs, on the other hand, did not benefit as much from the losses as the other approaches.
The structure loss (and the combined losses) did however further stabilize the training process since the transformation goals were explicitly incorporated into the loss function.

In our evaluation, the transformation quality was well reflected in the segmentation scores, but only partially for the classification.
It is particularly important to note that if the best approach is selected solely based on the resulting classification or segmentation scores, hallucination artifacts might be missed.
The artifacts can introduce a new type of bias to the images which could `accidentally' improve the segmentation or classification scores.
Finally, another factor that contributed to the mismatch of transformation quality and classification scores for the prostate biopsies is the inter-annotator variance.

Multiple factors play a role in deciding which of the three architectures should be used to perform bias transfer.
When a large number of domains are available, the training time can become a limiting factor.
Since U-Net cycleGANs only transform between two domains, every target domain requires a separate model and training.
Another common problem is a lack of data.
If few samples exist (e.g. $<100$ per domain), U-Net cycleGANs might have too many trainable parameters to learn an adequate transformation.
Both of these issues would warrant selecting FPGs instead since all domains are used to train a single generator.
The cycleGAN variations did not match the performance of U-Net cycleGAN and FPG.
Therefore, we do not recommend using vanilla cycleGANs for bias transfer in medicine.
Regarding the evaluation of generative approaches, no consensus on adequate evaluation metrics exists.
Here, the complementary metrics SSIM and FID were good indicators of hallucination artifacts and therefore suitable for evaluating bias transfer.

\subsection{Limitations}
The selection of the models and the additional losses in this paper focused on bias transfer for domains with small differences between them, i.e., a shift in staining.
Therefore, the results might not be reproducible for less similar domains.
Additionally, bias transfer has to be performed with caution if the domains have an inherent content bias.
A disease can become part of the `style' if one domain contains more healthy and the other more malformed samples~\cite{Cohen2018}.

\section{Conclusion and outlook}
The goal of this paper was to determine which deep generative approach results in the best bias transfer for medical images originating from IF confocal microscopy of kidney biopsies and H\&E stained microscopy images of prostate biopsies.
The performance on the test set mostly corroborated our findings obtained on the validation data.
U-Net cycleGAN with combined losses and FPG with structure loss had a stable training process and created hallucination artifact-free images that imitated the target domain, improving the segmentation or classification performance.
Our results show that bias transfer for histopathological images benefits from adding structure-based losses since they help with content preservation.
The combination of MS-SSIM loss and additional identity loss is especially helpful if bias transfer is performed for domains that only require small changes.
For medical image datasets with larger differences, e.g.\ a different staining type, the additional identity loss should only be used if it is weighted much lower than the MS-SSIM loss.
Furthermore, additional losses are not a universal solution.
The individual network architectures and approaches have to be adapted to the task at hand, as our differing results for the two modalities have shown.
In future projects, similar datasets from additional sites could be investigated to get a firmer grasp on which transformation goals can be covered by the selected additional losses and which further limitations might prevent using deep learning-based bias transfer for medical datasets.

\subsubsection*{Acknowledgements} This work was supported by DFG (SFB 1192 projects B8, B9 and C3) and by BMBF (eMed Consortia `Fibromap').

\bibliographystyle{splncs04}
\bibliography{main}

\begin{thebibliography}{10}
\providecommand{\url}[1]{\texttt{#1}}
\providecommand{\urlprefix}{URL }
\providecommand{\doi}[1]{https://doi.org/#1}

\bibitem{Armanious2019}
{Armanious}, K., {Tanwar}, A., {Abdulatif}, S., {Küstner}, T., {Gatidis}, S.,
  {Yang}, B.: Unsupervised adversarial correction of rigid mr motion artifacts.
  In: 2020 IEEE 17th International Symposium on Biomedical Imaging (ISBI). pp.
  1494--1498 (2020)

\bibitem{Arvaniti2018a}
Arvaniti, E., Fricker, K., Moret, M., Rupp, N., Hermanns, T., Fankhauser, C.,
  Wey, N., Wild, P., Rüschoff, J.H., Claassen, M.: {Replication Data for:
  Automated Gleason grading of prostate cancer tissue microarrays via deep
  learning} (2018)

\bibitem{Arvaniti2018b}
Arvaniti, E., Fricker, K.S., Moret, M., Rupp, N., Hermanns, T., Fankhauser, C.,
  Wey, N., Wild, P.J., Ruschoff, J.H., Claassen, M.: {Automated Gleason grading
  of prostate cancer tissue microarrays via deep learning}. Scientific Reports
  (2018)

\bibitem{DeBel2019}
de~Bel, T., Hermsen, M., {Jesper Kers}, R., van~der Laak, J., Litjens, G.:
  {Stain-Transforming Cycle-Consistent Generative Adversarial Networks for
  Improved Segmentation of Renal Histopathology}. In: Proceedings of The 2nd
  International Conference on Medical Imaging with Deep Learning. pp. 151--163
  (2019)

\bibitem{Chen2016}
Chen, N., Zhou, Q.: The evolving gleason grading system. Chinese journal of
  cancer research  \textbf{28}(1),  58--64 (2016)

\bibitem{Choi2018}
Choi, Y., Choi, M., Kim, M., Ha, J.W., Kim, S., Choo, J.: {StarGAN: Unified
  Generative Adversarial Networks for Multi-domain Image-to-Image Translation}.
  In: Proceedings of the IEEE Computer Society Conference on Computer Vision
  and Pattern Recognition. pp. 8789--8797 (2018)

\bibitem{Cohen2018}
Cohen, J.P., Luck, M., Honari, S.: {Distribution matching losses can
  hallucinate features in medical image translation}. In: Medical Image
  Computing and Computer Assisted Intervention -- MICCAI 2018. pp. 529--536
  (2018)

\bibitem{Deng2009}
Deng, J., Dong, W., Socher, R., Li, L.J., Li, K., Fei-Fei, L.: {ImageNet: A
  large-scale hierarchical image database}. In: 2009 IEEE Conference on
  Computer Vision and Pattern Recognition. pp. 248--255 (2009)

\bibitem{Dice1945}
Dice, L.R.: {Measures of the Amount of Ecologic Association Between Species}.
  Ecology  \textbf{26}(3),  297--302 (1945),
  \url{https://www.jstor.org/stable/1932409}

\bibitem{Egevad2013}
Egevad, L., Ahmad, A.S., Algaba, F., Berney, D.M., Boccon-Gibod, L.,
  Comp{\'{e}}rat, E., Evans, A.J., Griffiths, D., Grobholz, R., Kristiansen,
  G., Langner, C., Lopez-Beltran, A., Montironi, R., Moss, S., Oliveira, P.,
  Vainer, B., Varma, M., Camparo, P.: {Standardization of Gleason grading among
  337 European pathologists}. Histopathology  \textbf{62}(2),  247--256 (2013)

\bibitem{Engin2018}
Engin, D., Genc, A., Ekenel, H.K.: {Cycle-dehaze: Enhanced cyclegan for single
  image dehazing}. In: IEEE Computer Society Conference on Computer Vision and
  Pattern Recognition Workshops. pp. 938--946 (2018)

\bibitem{Gonzalez2007}
Gonzalez, R.C., Woods, R.E., Masters, B.R.: {Digital Image Processing, Third
  Edition}. Pearson (2007)

\bibitem{Heusel2017}
Heusel, M., Ramsauer, H., Unterthiner, T., Nessler, B., Hochreiter, S.: {GANs
  trained by a two time-scale update rule converge to a local Nash
  equilibrium}. In: Proceedings of the 31st International Conference on Neural
  Information Processing Systems. pp. 6629–--6640. NIPS'17 (2017)

\bibitem{Isola2017}
Isola, P., Zhu, J.Y., Zhou, T., Efros, A.A.: {Image-to-image translation with
  conditional adversarial networks}. In: Proceedings - 30th IEEE Conference on
  Computer Vision and Pattern Recognition, CVPR 2017. pp. 5967--5976 (2017)

\bibitem{Kohli2017}
Kohli, M.D., Summers, R.M., Geis, J.R.: {Medical Image Data and Datasets in the
  Era of Machine Learning - Whitepaper from the 2016 C-MIMI Meeting Dataset
  Session}. Journal of Digital Imaging  \textbf{30}(4),  392--399 (2017)

\bibitem{Ma2020}
Ma, Y., Liu, Y., Cheng, J., Zheng, Y., Ghahremani, M., Chen, H., Liu, J., Zhao,
  Y.: {Cycle Structure and Illumination Constrained GAN for Medical Image
  Enhancement}. In: Medical Image Computing and Computer Assisted Intervention
  -- MICCAI 2020. pp. 667--677 (2020)

\bibitem{Manakov2019}
Manakov, I., Rohm, M., Kern, C., Schworm, B., Kortuem, K., Tresp, V.: {Noise as
  Domain Shift: Denoising Medical Images by Unpaired Image Translation}. In:
  Domain Adaptation and Representation Transfer and Medical Image Learning with
  Less Labels and Imperfect Data. pp. 3--10 (2019)

\bibitem{Reinhard2001}
Reinhard, E., Ashikhmin, M., Gooch, B., Shirley, P.: {Color transfer between
  images}. IEEE Computer Graphics and Applications  \textbf{21}(5),  34--41
  (2001)

\bibitem{Ronneberger2015}
Ronneberger, O., Fischer, P., Brox, T.: {U-net: Convolutional networks for
  biomedical image segmentation}. In: Medical Image Computing and
  Computer-Assisted Intervention -- MICCAI 2015. pp. 234--241 (2015)

\bibitem{Shaban2019}
Shaban, M.T., Baur, C., Navab, N., Albarqouni, S.: {Staingan: Stain style
  transfer for digital histological images}. In: 2019 IEEE 16th International
  Symposium on Biomedical Imaging (ISBI 2019). pp. 953--956 (2019)

\bibitem{Siddiquee2019}
Siddiquee, M.M.R., Zhou, Z., Tajbakhsh, N., Feng, R., Gotway, M.B., Bengio, Y.,
  Liang, J.: {Learning Fixed Points in Generative Adversarial Networks: From
  Image-to-Image Translation to Disease Detection and Localization}. In: 2019
  IEEE/CVF International Conference on Computer Vision (ICCV). pp. 191--200
  (2019)

\bibitem{Wang2004}
Wang, Z., Bovik, A.C., Sheikh, H.R., Simoncelli, E.P.: {Image quality
  assessment: From error visibility to structural similarity}. IEEE
  Transactions on Image Processing  \textbf{13}(4),  600--612 (2004)

\bibitem{Wang2003}
Wang, Z., Simoncelli, E.P., Bovik, A.C.: {Multi-scale structural similarity for
  image quality assessment}. In: The Thirty-Seventh Asilomar Conference on
  Signals, Systems and Computers, 2003. pp. 1398--1402 (2003)

\bibitem{Zhu2017}
{Zhu}, J., {Park}, T., {Isola}, P., {Efros}, A.A.: Unpaired image-to-image
  translation using cycle-consistent adversarial networks. In: 2017 IEEE
  International Conference on Computer Vision (ICCV). pp. 2242--2251 (2017)

\bibitem{Zimmermann2021}
Zimmermann, M., Klaus, M., Wong, M., Thebille, A.K., Gernhold, L., Kuppe, C.,
  Halder, M., Kranz, J., Wanner, N., Braun, F., Wulf, S., Wiech, T., Panzer,
  U., Krebs, C., Hoxha, E., Kramann, R., Huber, T., Bonn, S., Puelles, V.: Deep
  learning-based molecular morphometrics for kidney biopsies. JCI Insight
  \textbf{6} (03 2021)

\end{thebibliography}
\end{document}